# Direct spectroscopic evidence for phase competition between the pseudogap and superconductivity in $Bi_2Sr_2CaCu_2O_{8+\delta}$

Makoto Hashimoto,[1] Elizabeth A. Nowadnick,[2,3] Rui-Hua. He,[2,3,4†] Inna M. Vishik,[2,3] Brian Moritz,[2,5] Yu He,[2,3] Kiyohisa Tanaka,[2,3,6] Robert G. Moore,[1,2] Donghui Lu,[1] Yoshiyuki Yoshida,[7] Motoyuki Ishikado,[7,8‡] Takao Sasagawa,[9] Kazuhiro Fujita,[10,11] Shigeyuki Ishida,[10] Shinichi Uchida,[10] Hiroshi Eisaki,[7] Zahid Hussain,[4] Thomas P. Devereaux,[2] and Zhi-Xun Shen[2,3*]

[1]*Stanford Synchrotron Radiation Lightsource, SLAC National Accelerator Laboratory, Menlo Park, California 94025, USA.*
[2]*Stanford Institute for Materials and Energy Sciences, SLAC National Accelerator Laboratory, Menlo Park, California 94025, USA.*
[3]*Geballe Laboratory for Advanced Materials, Departments of Physics and Applied Physics, Stanford University, Stanford, California 94305, USA.*
[4]*Advanced Light Source, Lawrence Berkeley National Laboratory, Berkeley, California 94720, USA.*
[5]*Department of Physics and Astrophysics, University of North Dakota, Grand Forks, North Dakota 58202, USA.*
[6]*Department of Physics, Osaka University, Toyonaka, Osaka 560-0043, Japan.*
[7]*National Institute of Advanced Industrial Science and Technology (AIST), Tsukuba, Ibaraki 305-8568, Japan.*
[8]*Quantum Beam Science Directorate, Japan Atomic Energy Agency, Tokai, Ibaraki 319-1195, Japan.*
[9]*Materials and Structures Laboratory, Tokyo institute of Technology, Yokohama, Kanagawa 226-8503, Japan.*
[10]*Department of Physics, University of Tokyo, Bunkyo-ku, Tokyo 113-0033, Japan.*
[11]*Laboratory for Atomic and Solid State Physics, Department of Physics, Cornell University, Ithaca, New York 14853, USA.*

[*]to whom correspondence should be addressed: zxshen@stanford.edu.
[†]Current addresses: *Department of Physics, Boston College, Chestnut Hill, MA 02467, USA.*
[‡]Current addresses: *Research Center for Neutron Science and Technology, Comprehensive Research Organization for Science and Society (CROSS), Tokai, Naka, Ibaraki 319-1106, Japan.*

**In the high-temperature ($T_c$) cuprate superconductors, increasing evidence suggests that the pseudogap[1], existing below the pseudogap temperature $T^*$, has a distinct broken electronic symmetry from that of superconductivity.[2-20] Particularly, recent scattering experiments on the underdoped cuprates have suggested that a charge ordering competes with superconductivity.[18-20] However, no direct link of this physics and the important low-energy excitations has been identified. Here we report an antagonistic singularity at $T_c$ in the spectral weight of $Bi_2Sr_2CaCu_2O_{8+\delta}$ as a compelling evidence for phase competition, which persists up to a high hole concentration $p \sim 0.22$. Comparison with a theoretical calculation confirms that the singularity is a signature of competition between the order parameters for the pseudogap and superconductivity. The observation of the spectroscopic singularity at finite temperatures over a wide doping range provides new insights into the nature of the competitive interplay between the two intertwined phases and the complex phase diagram near the pseudogap critical point.**

If the pseudogap and superconducting order parameters compete within a Ginzburg-Landau framework, this should be detectable as an abrupt change in the spectral-weight transfer at $T_c$. To search for this signature, we performed measurements of the electronic states in $Bi_2Sr_2CaCu_2O_{8+\delta}$ (Bi2212) using angle-resolved photoemission spectroscopy (ARPES), which directly probes the occupied states of the single-particle spectral function. ARPES is an ideal tool for this study because it can resolve the strong momentum anisotropies of the pseudogap and superconducting gap, both of which become the largest at the antinode, the Fermi momentum ($k_F$) on the Brillouin zone boundary (Fig. 1b).

We show in Fig. 1a a detailed temperature dependence of the ARPES spectra at the antinode of optimally-doped Bi2212 (denoted OP98, $p \sim 0.160$, $T_c = 98$ K). Here, all the spectra are divided by the resolution-convolved Fermi-Dirac function (FD) to effectively remove the Fermi cutoff. At $T \ll T_c$, the spectra show a "peak-dip-hump" structure which is typical for the cuprates near the antinode. While the peak (blue circles) is a signature of superconductivity, the dip (purple down triangles) and hump (red squares) are often associated with strong band renormalizations arising from electron-boson coupling.[21,22] Above $T_c$, the spectra show a continued suppression of spectral intensity at the Fermi level ($E_F$), defining the pseudogap.[1] Notably, the peak feature becomes weaker but survives above $T_c$. There is no singular signature in the spectral lineshape at $T_c$ over a wide doping range (Supplementary Fig. 1 for complete dataset). The non-trivial evolution of the spectral lineshape has been making the interpretation of the pseudogap difficult.

To investigate the nature of the peak, dip and hump, we show in Fig. 1c their energies as a function of temperature. The energy scale of the anomalously broad hump feature at $T_c < T < T^*$ decreases with increasing temperature and hole doping (Supplementary Fig. 1), suggesting that it arises from the pseudogap. The hump at $T > T_c$ continuously connects with that at $T < T_c$ (Fig. 1c), suggesting that not only the electron-boson coupling but also the pseudogap affects the hump energy at $T < T_c$ while simultaneously coexisting with the superconducting peak. Here, a simple addition of two gaps in quadrature does not reproduce the data and does not capture the mixed nature of the all spectral features as noted earlier.[15]

Next, we show in Figs. 1d-1f the spectral weight obtained by analyzing the spectral intensity $I(\omega)$ at the antinode (Fig. 1a), where $\omega$ is energy. Fig. 1d shows the 1$^{st}$ moment defined as $\int_{0eV}^{0.25eV} \omega I(\omega) d\omega \Big/ \int_{0eV}^{0.25eV} I(\omega) d\omega$, which gives the spectral-weight center of mass. Figs. 1e and

1F show spectral weights in the ranges [0, 0.07] and [0.20, 0.25] eV, which we denote by low- and high-energy spectral weights, respectively Because the energy scale for superconductivity is < 50 meV, the opening of a superconducting gap at $k_F$ should push the 1$^{st}$ moment energy away from $E_F$ in a narrow range, and have almost no effect on the low- and high-energy spectral weights.

In contrast with the behavior expected for homogeneous superconductivity, the most striking signature in the current result is the spectral-weight singularity at $T_c$ (Figs. 1d-1f). The spectral weight is clearly sensitive to the sharp onset of coherent superconductivity at $T_c$, different from the spectral lineshape (Fig. S1). For $T > T_c$, the 1$^{st}$ moment moves to a higher energy with decreasing temperature, while the low/high-energy spectral weight decreases/increases, respectively. This suggests that the opening of the pseudogap involves spectral weight reorganization over a wide energy range greater than a few hundred meV. Upon lowering the temperature to $T < T_c$, the 1$^{st}$ moment and the low- and high-energy spectral weights all show the opposite trends from those at $T > T_c$. This "antagonistic" singularity suggests that the pseudogap spectral weight at higher energies participates in forming the coherent superconducting peak that emerges below $T_c$.

To further understand this result, we have considered a competition between the pseudogap and superconducting order parameters within a Ginzburg-Landau treatment (Supplementary Information for calculation detail). We show in Fig. 2a the temperature dependence of the calculated antinodal spectra across $T_c$. Here, the pseudogap order parameter, which we model as fluctuating density-wave order, starts to decrease once the superconducting order parameter becomes non-zero below $T_c$ (Fig. 2c). The anomalously broad spectral lineshape in the pseudogap state and the spectral weight redistribution over the wide energy range are captured by inputting a short correlation length of the pseudogap order. The temperature dependence of the 1st moment and the low- and high-energy spectral weights in the calculation (open markers in Figs. 2e-2g, respectively) shows a singularity at $T_c$, consistent with the ARPES results (Fig. 1d-1f). Although the exact form of the pseudogap needs to be further investigated, this phenomenology strongly supports the picture that the pseudogap and superconducting orders compete in the superconducting state, and therefore $T_c$ is suppressed by the pseudogap order.

Our result may be reconciled with other observations of the existence of a distinct order. A similar spectral weight transfer between low- and high-energies with a clear signature at $T_c$ has been observed in the c-axis conductivity of underdoped $NdBa_2Cu_3O_{6.9}$.[23] In support of our choice of model for the pseudogap, we note that a density-wave-like order with short coherence length that breaks translational symmetry may be consistent with the antinodal dispersion observed in the pseudogap state of $Bi_{1.5}Pb_{0.55}Sr_{1.6}La_{0.4}CuO_{6+\delta}$,[12,15] and translational symmetry breaking observed at the surface[3,5,10,11] and in the bulk states in the underdoped regime.[17-20] In particular, the characteristic wave vectors for the charge ordering have been observed recently in the Bi-based cuprates, which is consistent between scattering and STM measurements[19,20] and suppressed below $T_c$.[19] The spectral-weight singularity at $T_c$ in the present result suggests a close relationship between the charge ordering and the ARPES pseudogap in the antinodal region, and provides information about its momentum structure. However, the observation of the charge modulation in the bulk states has been limited in a narrow doing range in the underdoped regime so far, and whether the charge ordering observed by scattering experiments is directly tied to the pseudogap physics remains an open question. Further exploration is needed to understand the relationship between the present result, other symmetry breaking[2,4,6-9,13,14,16] and the Fermi surface reconstruction.[24]

Another important piece of the antinodal physics is the electron-boson coupling, which may manifest itself as the dip and hump.[21,22] As shown in Fig. 1c, the hump is pushed further away from $E_F$ below $T_c$. This temperature dependence cannot be modeled if one considers only a competition between the pseudogap and superconductivity (blue open circles in Fig. 2d). To address this discrepancy, we consider a coupling of electrons to the $B_{1g}$ 'buckling' phonons (35 meV),[22] in combination with superconductivity and the pseudogap. This qualitatively reproduces the anomalous temperature dependence of the hump across $T_c$ (filled red squares in Fig. 2d), in addition to producing singularities at $T_c$ in the 1$^{st}$ moment and the low- and high-energy spectral weights (filled markers in Figs. 2e-2g, respectively). The result supports the idea that electron-boson coupling is essential to understand the antinodal spectral lineshape and its spectroscopic signature is intimately tied to the underlying quantum phases. Further, the simulation suggests that the suppression of the pseudogap below $T_c$ is key for understanding the emergence of the superconducting peak and the enhancement of the peak-dip-hump structure (Supplementary Information).

Figure 3 summarizes the temperature dependence of the spectral weight for underdoped $p = 0.132$ to overdoped $p = 0.224$ samples. As the doping increases, the 1$^{st}$ moment moves towards lower energies and low/high-energy spectral weights increase/decrease, similar to the $B_{1g}$ Raman spectra of $La_{2-x}Sr_xCuO_4$ and other cuprates.[25] The spectral-weight singularity at $T_c$ persists up to $p = 0.207$ (OD80, Fig. 3d) and possibly $p = 0.218$ (OD71, Fig. 3e). However, at $p = 0.224$ (OD65, Fig. 3f), the singularity becomes undetectable, and the result can be understood by the opening of a superconducting gap alone with a diminished or absent pseudogap. The slight decrease/increase of the low/high-energy spectral weight below $T_c$ may be due to the effects of electron-boson coupling and/or the tail of the superconducting peak.

The increase in the low-energy spectral weight below $T_c$ (dashed arrow in Fig. 1e for OP98) provides us with an estimate of how much pseudogap spectral weight at higher energies contributes to the superconducting peak, and its doping dependence is plotted in Fig. 4a. This increase rapidly becomes smaller with doping and non-detectable at $p = 0.224$. This suggests that the competition between the pseudogap and superconducting orders becomes weaker with doping, but the pseudogap as a competing order persists at finite temperatures and coexists with superconductivity below $T_c$ up to at least $p \sim 0.22$. We emphasize that the spectral weigh analysis enabled us to detect such a clear feature for the competition and its disappearance at finite temperatures especially in the overdoped regime, which cannot be easily addressed by the spectral lineshape analysis.

We show in Fig. 4b the antinodal spectra at $T \ll T_c$ for different dopings, and in Fig. 4c their peak, dip and hump energies as a function of doping. At $p > 0.19$, all the energies show similar doping dependence. The constant offset of -35 meV for the dip energy can be understood as arising from the mode energy of ~35 meV. In contrast, at $p < 0.19$, the peak, dip and hump energies do not show strong doping dependences. These abrupt changes at $p \sim 0.19$ in the ground state suggest a potential critical point, which could be consistent with the recently reported doping dependence of the near-nodal gap slope [26] ($v_\Delta$ in Fig. 4c) and the anomalies observed by ARPES,[27,28] NMR,[29] ultrasound,[16] and transport[30,31] studies. Moreover, a signature of the pseudogap at finite temperature is observed at higher doping levels than the ground state pseudogap critical point at $p \sim 0.19$. This is consistent with and solidifies the recently proposed phase diagram,[26,32] where that the pseudogap line bends back in the superconducting dome due to phase competition, separating the coexisting/competing phase and the purely superconducting phase. Finally, the most striking finding in Figs. 4b and c is the sudden jump of the hump energy across $p \sim 0.19$. This doping dependent hump energy discontinuity suggests that the pseudogap effect on the hump comes into play only at $p < 0.19$ for low temperatures. This result strongly supports our interpretation gained from the temperature dependence (Fig. 1) and the simulation (Fig. 2) that the hump is strongly affected by the pseudogap.

We showed an unforeseen antagonistic singularity at $T_c$ in the spectral weight as a direct spectroscopic evidence for the competition between the pseudogap and superconducting orders. The observation of such a signature for the competition over a wide doping range suggests that the phase competition recently observed in the underdoped regime by scattering experiments[18-20] may not be limited to the underdoped regime, but extends to the overdoped regime as a universal and intrinsic aspect of the interplay between the pseudogap and superconductivity. Further scattering studies are desired to pin down the exact relationship between the charge ordering and the pseudogap. Because such a clear onset at $T_c$ is not present in the spectral lineshape (Supplementary Fig. 1), our finding of the singularity in the low-energy spectral weight provides a crucial piece of information about the nature of the pseudogap and the competitive interplay between the pseudogap and superconductivity, providing us with a foundation for a holistic understanding of the phase diagram and mechanism of high-$T_c$.

**Methods**

*Samples* High-quality single crystals of $Bi_{1.54}Pb_{0.6}Sr_{1.88}CaCu_2O_{8+\delta}$, $Bi_2Sr_2CaCu_2O_{8+\delta}$, $Bi_{2-x}Sr_{2+x}CaCu_2O_{8+\delta}$, and $Bi_2Sr_2(Ca,Dy)Cu_2O_{8+\delta}$, which are various families of $Bi_2Sr_2CaCu_2O_{8+\delta}$ (Bi2212), were grown by the floating-zone method. The hole concentration $p$ was controlled by annealing the samples in $N_2$ or $O_2$ flow. Detailed experimental conditions for each sample are summarized in Table. S1. We determined $p$ from $T_c$ via an empirical curve, $T_c = T_{c,max}[1-82.6(p-0.16)^2]$, taking 98K as the optimum $T_c$ for Bi2212.[33] Pb doping significantly suppresses the BiO superlattice modulation, which allows us to discuss more quantitatively the electronic structure, particularly around the antinode.[12,15]

*Measurements*   ARPES measurements were performed at beamline 5-4 of SSRL, SLAC National Accelerator Laboratory using a Scienta R4000 electron analyzer. Photon energy was tuned to highlight either the antibonding band or bonding band at the antinode.[34] In the main text, focus is mostly placed on the antibonding band taken with photon energy of 18.4 eV. The photon polarization was fixed parallel to the Cu-O bonding direction and perpendicular to the measured cuts. The energy resolution and the angular resolution along the analyzer slit were set at ~10 meV and ~0.13°, respectively. The samples were cleaved *in-situ* and the sample temperature was varied from 7 K to 240 K. The vacuum was kept better than $4.0 \times 10^{-11}$ Torr throughout the measurement. $E_F$ was calibrated using a gold sample electronically connected to the measured sample. Measurements were performed on several samples with consistent results.

**Acknowledgments**

We thank fruitful discussion with S. Kivelson, H. Yao, A. Millis, D. Scalapino, P. Hirshfeld, B. Markiewicz, D.-H. Lee, L. Yu, A. Fujimori & N. Nagaosa.  ARPES experiments were performed at the Stanford Synchrotron Radiation Lightsource, operated by the Office of Basic Energy Science, U.S. DOE. This work is supported by DOE Office of Basic Energy Sciences, Materials Sciences and Engineering Division, under Contract DE-AC02-76SF00515.

**Figure Legends**

**Fig. 1**. Temperature dependence of the antinodal electronic states in optimally-doped Bi2212. **a** FD-divided spectra at $AN^+$. **b** Schematic Fermi surface. Two antinodal momenta for the antibonding band, $AN^-$ and $AN^+$, are indicated. **c** Peak, dip, and hump energies as a function of temperature. **d-f** 1$^{st}$ moment in the range [0, 0.25] eV, the spectral weights in the ranges [0, 0.07] and [0.20, 0.25] eV, respectively. Error bars are estimated to be smaller than the symbol sizes. The dashed arrow in panel e indicates the increase in the low-energy spectral weight below $T_c$. For the detail of the spectral weight analysis, see Supplementary Information.

**Fig. 2.** Simulated temperature dependence of the antinodal spectra with the pseudogap, electron-phonon coupling, and superconductivity. **a,b** Spectra across $T_c$ without and with coupling to 35 meV phonons, respectively. **c** Pseudogap (PG) and superconducting (SC) order parameters. **d** Hump and peak energies. **e-g** 1$^{st}$ moment in the range [0, 0.20] eV, spectral weights in the ranges [0, 0.07] and [0.15, 0.20] eV, respectively. See Supplementary Information for the detail of the calculation.

**Fig. 3.** Doping dependence of competition between the pseudogap and superconductivity. **a-f** Antinodal spectral weights in the ranges [0, 0.07] and [0.20, 0.25] eV (left axis) and the 1$^{st}$ moment in the range [0, 0.25] eV (right axis) for underdoped to overdoped Bi2212. Error bars are estimated to be smaller than the symbol sizes. $T_c$ and $T_{AN}$, the temperature at which the antinodal gap closes (Supplementary Fig. 1), are indicated by dashed lines.

**Fig. 4** Pseudogap critical point in Bi2212. **a** Magnitude of the increase in the low-energy spectral weight below $T_c$ (see main text and Fig. 1e) as a function of doping as an estimate . The magnitudes are normalized to those at $p = 0.132$. Error bars are estimated to be ±0.1. **b** Doping dependence of the antinodal spectra at $T \ll T_c$. **c** Peak, dip, and hump energies as a function of doping. The black arrow indicates the possible pseudogap critical point in the ground state at

$p=0.19\pm0.01$. The near-nodal gap slope $v_\Delta$ is reproduced from Ref. [26]. Red and black dashed lines are guide for eyes.

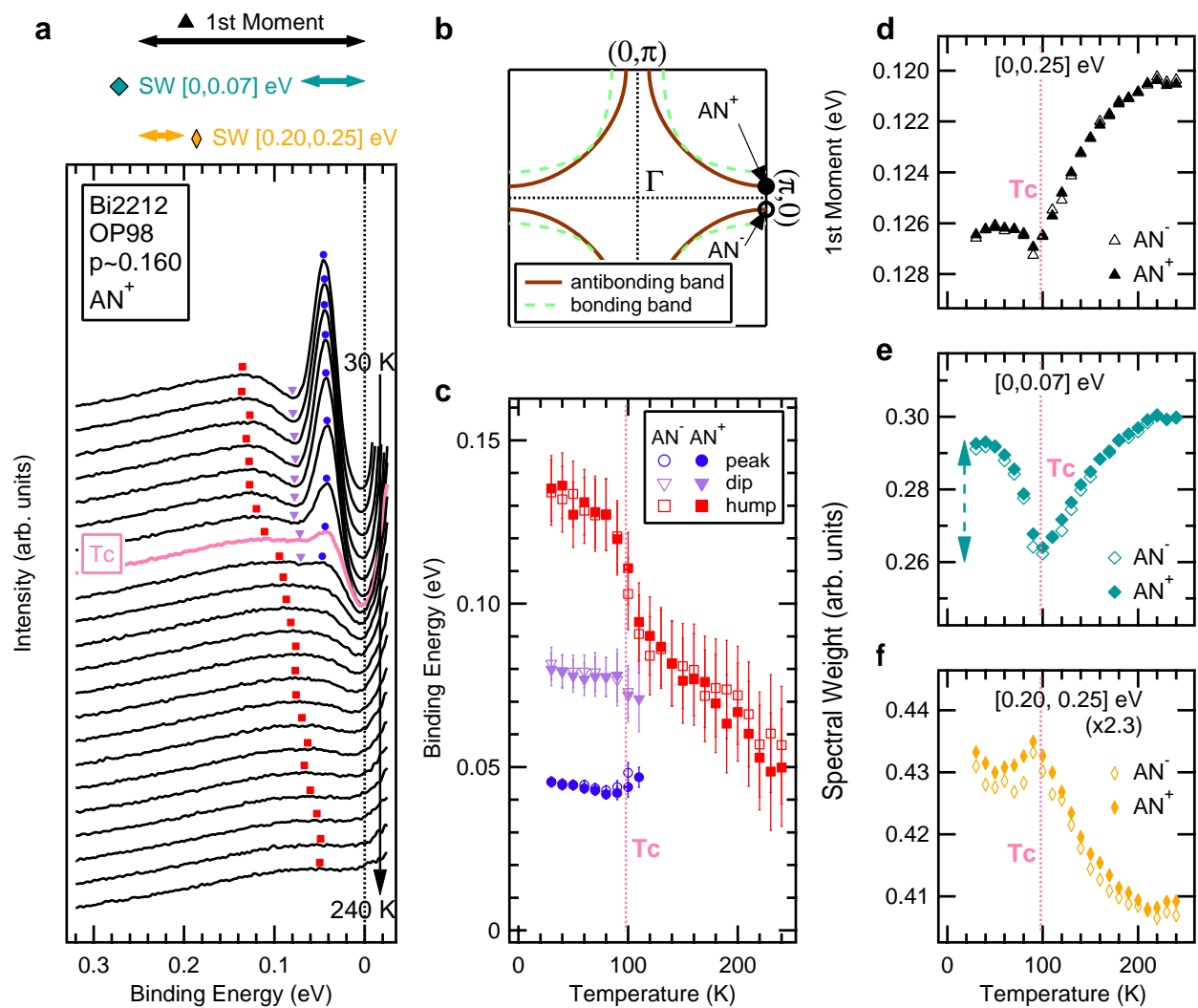

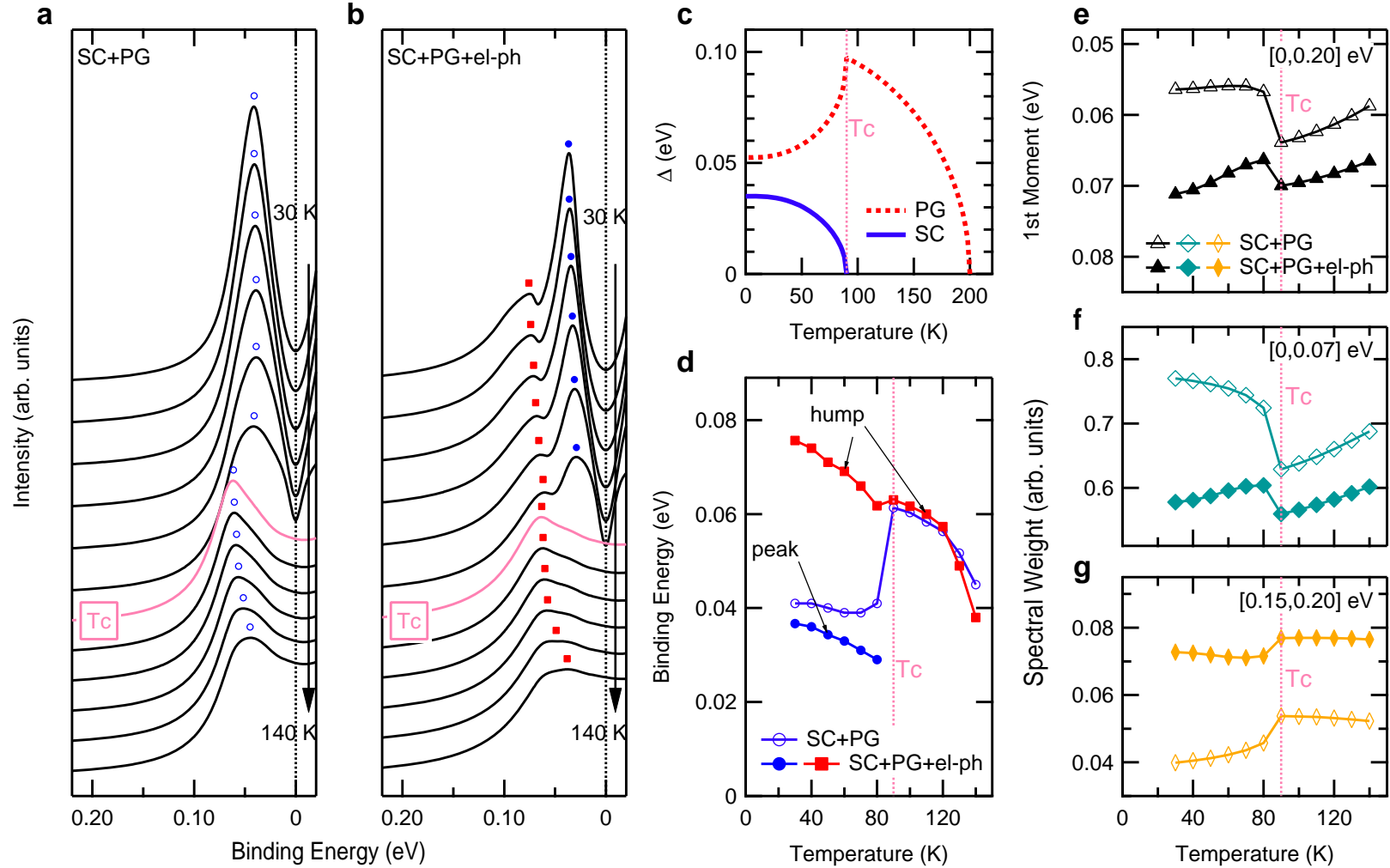

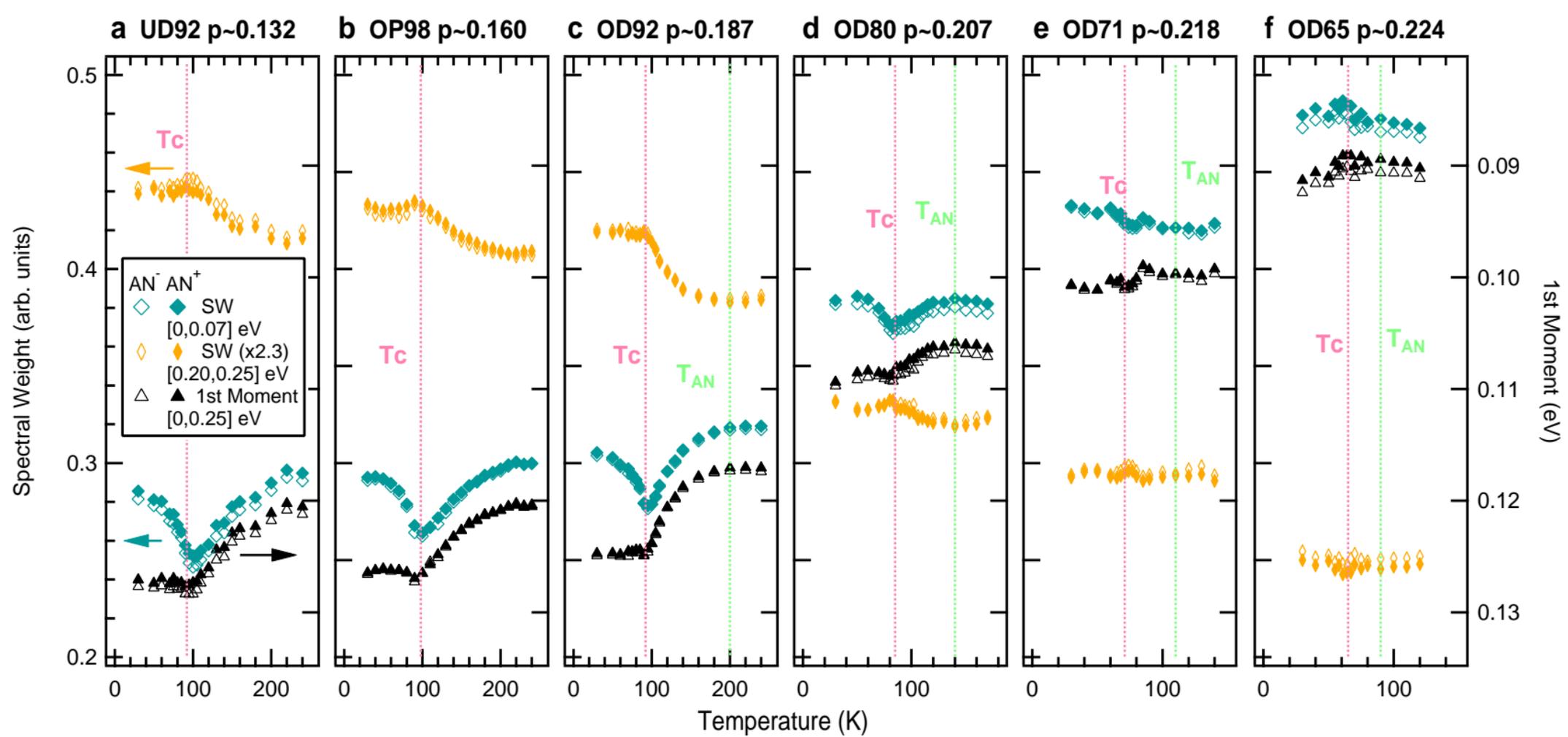

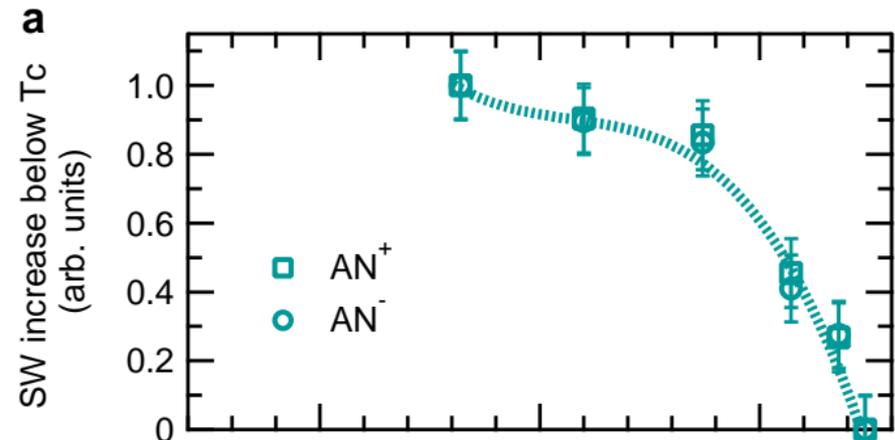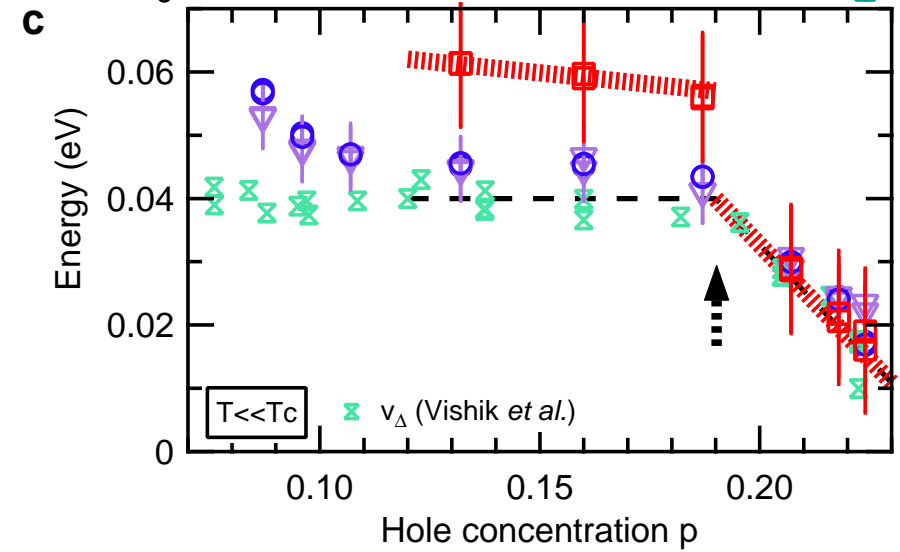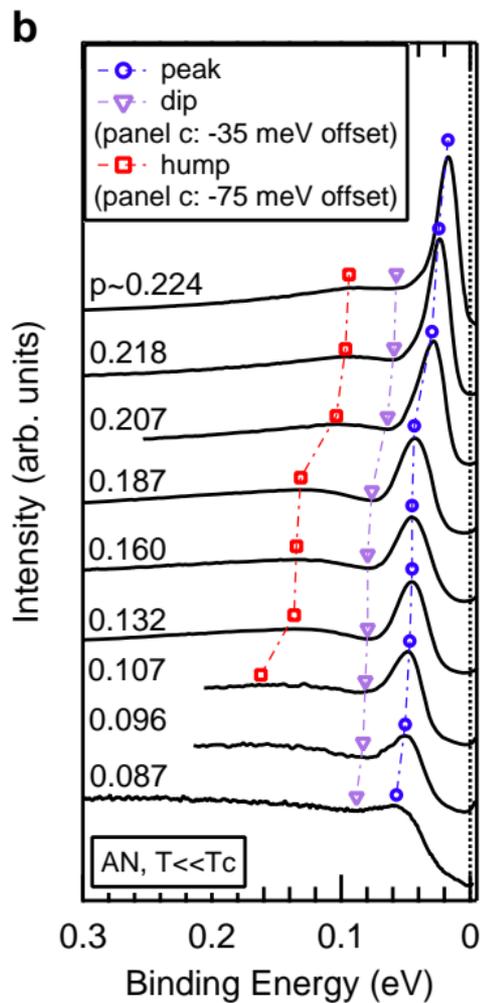

# Supplementary Information for Direct spectroscopic evidence for phase competition between the pseudogap and superconductivity in $Bi_2Sr_2CaCu_2O_{8+\delta}$

Makoto Hashimoto,[1] Elizabeth A. Nowadnick,[2,3] Rui-Hua. He,[2,3,4†] Inna M. Vishik,[2,3] Brian Moritz,[2,5] Yu He,[2,3] Kiyohisa Tanaka,[2,3,6] Robert G. Moore,[1,2] Donghui Lu,[1] Yoshiyuki Yoshida,[7] Motoyuki Ishikado,[7,8‡] Takao Sasagawa,[9] Kazuhiro Fujita,[10,11] Shigeyuki Ishida,[10] Shinichi Uchida,[10] Hiroshi Eisaki,[7] Zahid Hussain,[4] Thomas P. Devereaux,[2] and Zhi-Xun Shen[2,3*]

[1]*Stanford Synchrotron Radiation Lightsource, SLAC National Accelerator Laboratory, Menlo Park, California 94025, USA.*
[2]*Stanford Institute for Materials and Energy Sciences, SLAC National Accelerator Laboratory, Menlo Park, California 94025, USA.*
[3]*Geballe Laboratory for Advanced Materials, Departments of Physics and Applied Physics, Stanford University, Stanford, California 94305, USA.*
[4]*Advanced Light Source, Lawrence Berkeley National Laboratory, Berkeley, California 94720, USA.*
[5]*Department of Physics and Astrophysics, University of North Dakota, Grand Forks, North Dakota 58202, USA.*
[6]*Department of Physics, Osaka University, Toyonaka, Osaka 560-0043, Japan.*
[7]*National Institute of Advanced Industrial Science and Technology (AIST), Tsukuba, Ibaraki 305-8568, Japan.*
[8]*Quantum Beam Science Directorate, Japan Atomic Energy Agency, Tokai, Ibaraki 319-1195, Japan.*
[9]*Materials and Structures Laboratory, Tokyo institute of Technology, Yokohama, Kanagawa 226-8503, Japan.*
[10]*Department of Physics, University of Tokyo, Bunkyo-ku, Tokyo 113-0033, Japan.*
[11]*Laboratory for Atomic and Solid State Physics, Department of Physics, Cornell University, Ithaca, New York 14853, USA.*

[*]to whom correspondence should be addressed: zxshen@stanford.edu.
[†]Current addresses: *Department of Physics, Boston College, Chestnut Hill, MA 02467, USA.*
[‡]Current addresses: *Research Center for Neutron Science and Technology, Comprehensive Research Organization for Science and Society (CROSS), Tokai, Naka, Ibaraki 319-1106, Japan.*

**Table S1. Samples and experimental conditions**

| Sample | $T_c$ (K) | Composition | Measurement Temperature (K) | Photon energy (eV) |
|---|---|---|---|---|
| OD65 | 65 | $Bi_{1.54}Pb_{0.6}Sr_{1.88}CaCu_2O_{8+\delta}$ | 30 – 140 | 18.4 |
| OD71 | 71 | $Bi_{1.54}Pb_{0.6}Sr_{1.88}CaCu_2O_{8+\delta}$ | 30 – 140 | 18.4 |
| OD80 | 80 | $Bi_{1.54}Pb_{0.6}Sr_{1.88}CaCu_2O_{8+\delta}$ | 30 – 170 | 18.4 |
| OD92 | 92 | $Bi_{1.54}Pb_{0.6}Sr_{1.88}CaCu_2O_{8+\delta}$ | 30 – 240 | 18.4 |
| OP98 | 98 | $Bi_{1.54}Pb_{0.6}Sr_{1.88}CaCu_2O_{8+\delta}$ | 30 – 240 | 18.4 |
| OP98 | 98 | $Bi_{1.54}Pb_{0.6}Sr_{1.88}CaCu_2O_{8+\delta}$ | 30 – 140 | 21.0 (Fig. S3) |
| UD92 | 92 | $Bi_{1.54}Pb_{0.6}Sr_{1.88}CaCu_2O_{8+\delta}$ | 30 – 240 | 18.4 |
| UD75 | 75 | $Bi_2Sr_2CaCu_2O_{8+\delta}$ | 10 | 22.7 |
| UD65 | 65 | $Bi_{2-x}Sr_{2+x}CaCu_2O_{8+\delta}$ | 12 | 22.7 |
| UD55 | 55 | $Bi_2Sr_2(Ca,Dy)Cu_2O_{8+\delta}$ | 7 | 21.0 |

*Fermi momentum $k_F$*

For OD65, OD71, OD85 and OD92, $T^*$ is within the measurement temperature window (Table S1). Above $T^*$, the peak position of the FD-divided spectra shows a parabolic dispersion. $k_F$ is defined as the Femi crossing of the dispersion, similar to previous reports.[1,2] For other samples, $T^*$ is out of the measurement temperature window and $k_F$ is defined as the momentum where the momentum distribution curve (MDC) at $E_F$ shows a peak at temperatures closest to $T^*$, which has been shown to give a good estimate of $k_F$[1,3,4] and often used to determine $k_F$ in ARPES studies. We have checked that small differences in $k_F$ estimation do not affect our conclusion. Note that the detector nonlinearity of the Scienta electron analyzer[5-7] is calibrated and corrected. The summary of the doping and temperature dependence of the antinodal spectra in Bi2212 is shown in Fig. S1.

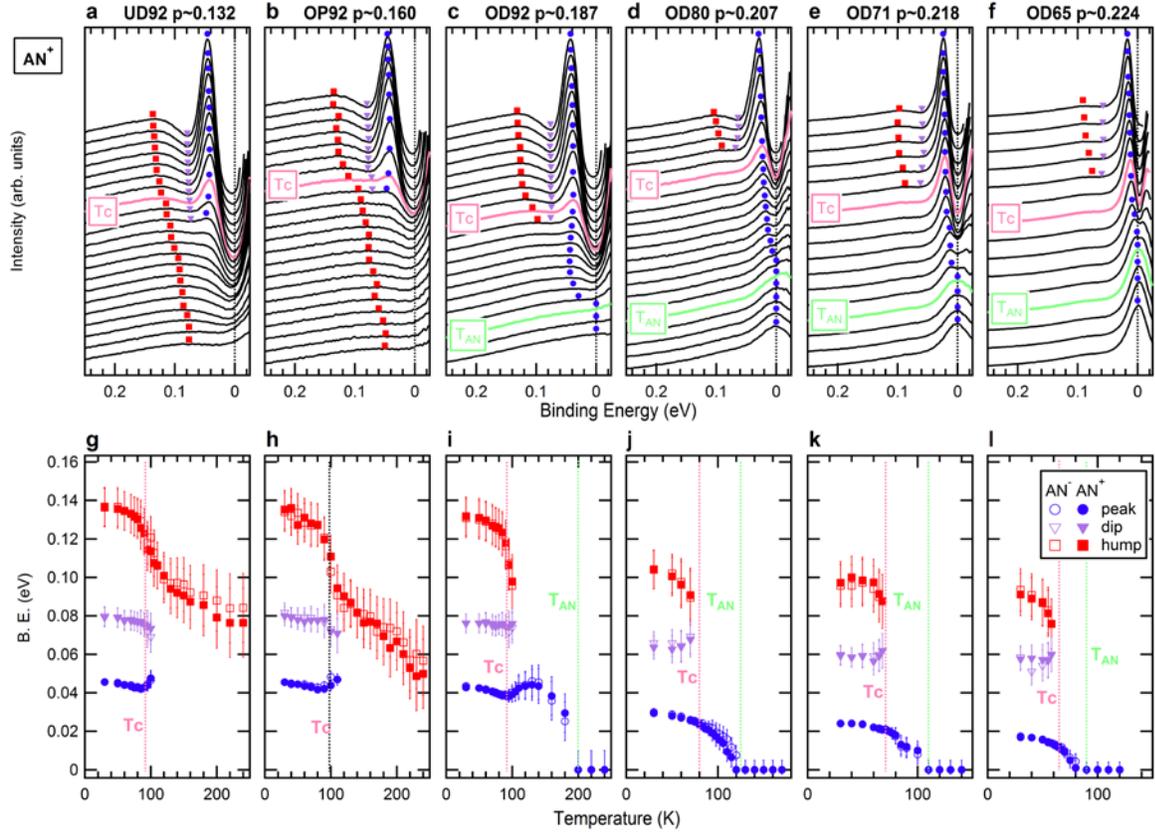

**Fig. S1. Doping and temperature dependence of the antinodal spectra in Bi2212. a-f** Temperature dependence of the FD-divided spectra at $AN^+$. Peak, dip, and hump features are indicated by markers. **g-l** Peak, dip, and hump energies as a function of temperature. Panels b and h are reproduced from main text Fig. 1a and 1c. $T_{AN}$ is the temperature where the spectra show a peak at $E_F$.

*Spectral weight analysis*

Figure S2a shows the temperature dependence of the low-energy spectral weight obtained from two different normalizations of the ARPES intensity. The spectral weights in red circles are normalized to the intensity at energies well above $E_F$ where the spectra have background signal from the synchrotron light of higher orders. Because the intensity ratio between the higher order light and the main first order light is stable and constant with time owing to the stable synchrotron operation in top-off (top-up) mode, this procedure effectively normalizes the intensity to the photon flux at the sample with high accuracy. We have checked that the normalization of the ARPES intensity to the photocurrent measured at a refocusing mirror just before the ARPES endstation shows consistent results. The good agreement between the normalization to the intensity well above $E_F$ (red) and to the total spectral weight (blue) $\int_{0\text{eV}}^{0.07\text{eV}} I(\omega)d\omega \Big/ \int_{0\text{eV}}^{0.25\text{eV}} I(\omega)d\omega$, which are reproduced from main text Fig. 1, ensures that our conclusions do not change with normalization of the ARPES intensity. The temperature dependence of the high-energy spectral weight also shows consistent results between the two normalizations (Fig. S2b). In addition, the 1$^{st}$ moment is complementary analysis to the low- and high-energy spectral weights because it is independent from the normalization of the ARPES intensity. The consistency between the spectral weights and 1$^{st}$ moment also supports our conclusions. In Figs. S2c and S2d, we show the energy-window dependence of the spectral weight near $E_F$ and at higher energies, respectively. The spectral-weight singularity at $T_c$ is clearly visible over different energy windows, confirming that our conclusion is robust against the integration energy window.

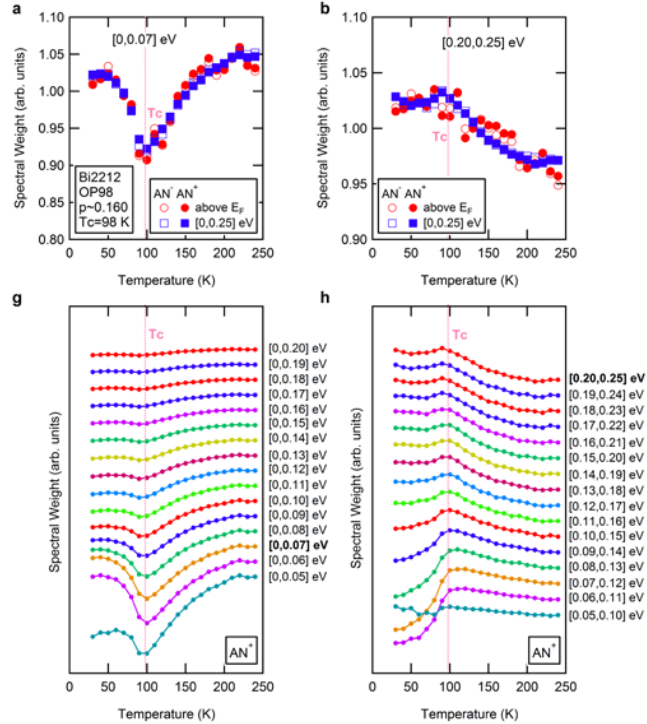

**Fig. S2. Normalization and integration-window dependences of the spectral weight in OP98. a,b** Normalization dependence of the low- and high-energy spectral weights in the ranges [0, 0.07] and [0.20, 0.25] eV, respectively. Red dots: spectral weight at each temperature is normalized to the total spectral weight over [0, 0.25] eV. Blue squares: spectral weight at each temperature is normalized to the background intensity well above $E_F$, which is proportional to the incident photon intensity on the sample surface. Obtained curves are normalized to the total area [30, 240] K for comparison. **c,d** Integration window dependence of the low- and high-energy spectral weights, respectively. Spectral weight at each temperature is normalized to the total spectral weight in the range [-0.25, 0] eV. Obtained curves are normalized to the spectral weight at 240 K for comparison.

*Bonding band*

The data shown in the main text Figs. 1-3 are for the antibonding band measured with 18.4 eV photons. We chose to focus on the antibonding band because the antibonding band can be well separated from the bonding band in the ARPES signal at 18.4 eV. At $T > T^*$, the MDC at $E_F$ does not show any signature of the bonding band in the underdoped regime. We estimate from fitting to four Lorentzians that the intensity for the bonding band is much less than 1 % of the total intensity for the underdoped and optimally-doped samples. With hole doping, the bonding band signal becomes gradually stronger and becomes ~3 % of the total intensity for OD65 Although the effect of the antibonding band may not be negligible in the ARPES spectra measured with 21 eV photons, one may be able to discuss the antinodal $k_F$ spectra for the bonding band because the antibonding band at $T > T^*$ is located above $E_F$. We show in Fig. S3 the ARPES spectra highlighting the bonding band at the antinode for OP98 by choosing a proper photon energy (21 eV).[8] The temperature dependence of the peak, dip, and hump energies at the bonding band antinode (indicated in Fig. S4b) are shown in Fig. S4c, which will be discussed in the next section. The temperature dependences of the 1$^{st}$ moment, low- and high-energy spectral weights are plotted in Fig. 4d. Consistent with the antibonding band (Fig. 1 in the main text), the temperature dependence of the low- and high-energy spectral weights and the 1$^{st}$ moment show an abrupt sign change at $T_c$, indicating that our conclusion holds not only for the antibonding band but also for the bonding band. In addition, similar temperature dependences have been observed for the spectral lineshape and the spectral weight at momenta between the two antinodal $k_F$'s both for the bonding band (21 eV) and antibonding band (18.4 eV), suggesting that the entire spectral function shows a consistent temperature dependence and the separation of the bonding band and the antibonding band does not affect our conclusion.

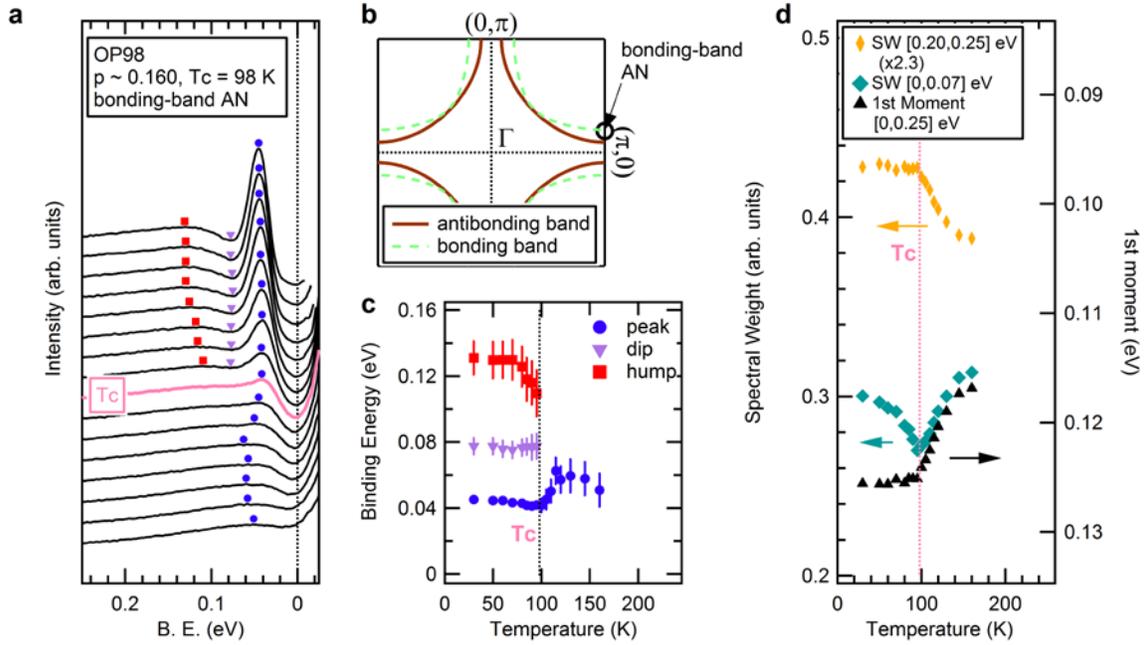

**Fig. S3. Temperature dependence of the ARPES spectra at the bonding-band antinode in OP98 taken with 21 eV photon energy. a** FD-divided spectra at bonding-band AN across $T_c$. Peak, dip, and hump features are indicated by markers. **b** Schematic Fermi surface. Bonding-band antinode (AN) is indicated. **c** Peak, dip, and hump energies as a function of temperature. **d** Temperature dependence of the 1st moment in the range [0, 0.25] eV, the spectral weights in the ranges [0, 0.07] and [0.20, 0.25] eV. Error bars are estimated to be smaller than the symbol sizes.

*Emergence of the superconducting peak below $T_c$*

We showed in the main text that the pseudogap, superconductivity, and electron-boson coupling all leave distinct fingerprints in the spectral lineshape. Our observation of competition between the pseudogap and superconducting order parameters provides us with an explanation for the long standing puzzle in the emergence of the superconducting peak at the antinode. The peak (blue dots in main text Fig. 1a) emerges below ~$T_c$ without well-defined quasi-particle peak above $T_c$ in the underdoped and optimally-doped regime following the temperature dependence of the superfluid density, consistent with previous reports [9,10]. This suggests that the peak is closely related with superconductivity in an unconventional manner, because in conventional superconductors, a well-defined quasi-particle peak exists above $T_c$. The simulated spectra (main text Fig. 2b) demonstrate how the superconducting peak below ~$T_c$ can emerge in the underdoped and optimally-doped regime. As suggested by the temperature dependence of the spectral weight (main text Fig. 3), coherent superconductivity transfers high-energy spectral weight for the pseudogap to low energy, dramatically enhancing the superconducting peak. This consequently emphasizes the dip by a pile-up of the spectral intensity around the superconducting peak energy,[11] which becomes available for electron-boson couplings below $T_c$. This suggests that the sharp bosonic mode does not need to form at ~$T_c$, but rather it can exist above $T_c$. The competition between the pseudogap and superconducting order parameters becomes key to understanding the emergence and enhancement of the peak-dip-hump structure in the superconducting state.

Although the pseudogap (hump) and superconducting (peak) features are well separated in the spectral lineshape at the antinode for the antibonding band in the underdoped and optimally-doped regimes (Fig. 1 and Fig. S1), they are not well distinguished around $T_c$ in the overdoped regime because of their closer energy scales (Fig. S1). In the overdoped regime, slightly above $T_c$, the intensity for the superconducting feature rapidly becomes small, and the dominant character of the peak is gradually taken over by the pseudogap energy feature that has similar energy scale to the superconductivity with increasing temperature. This may explain the peculiar temperature dependence of the peak for OD92 (Fig. S1i). The peak energy becomes smaller with increasing temperature at $T < T_c$, shows the opposite temperature dependence at $T_c < T < 130$ K, and becomes smaller again at $T > 130$ K until it becomes a quati-particle peak for the states at $T > T^*$. This non-monotonic temperature dependence of the peak cannot be explained by a single order parameter. Rather, the pseudogap and superconducting features slightly above $T_c$ are strongly mixed in the spectral lineshape, whereas the peak at $T < T_c$ and that at 130 K $< T < T^*$ may be dominantly associated with superconductivity and the pseudogap, respectively. This interpretation is supported by the coexistence of the pseudogap and superconductivity in the overdoped regime ($p < 0.22$), which is suggested by the spectral weight analysis in the main text.

This picture may also explain the contrast between the temperature dependences of the bonding and antibonding band spectra for OP98. While the hump survives at $T > T_c$ for the antibonding band (Fig. 1a), it disappears at $T > T_c$ for the bonding band (Fig. S3a) with a non-monotonic temperature dependence of the hump (Fig. S3b) similar to OD92 antibonding band (Fig. S1i). This suggests that the hump and peak features coexist slightly above $T_c$ both for the both bonding and antibonding bands, but cannot be disentangled in the spectral lineshape for the bonding band because the two energies are closer than those for the antibonding band. This can happen if the pseudogap is density-wave-like gap which is discussed in the main text because it opens the pseudogap in a momentum dependent manner.

We show in Fig. S4a the spectra above and well below $T_c$ for UD92 as a typical experimental example for the phenomenology described above. Here, at $T > T_c$, the peak disappears and hump is the main spectral feature. In contrast, for a more overdoped sample OD80 (Fig. S4b), the superconducting peak is smoothly connected with the quasi-particle peak above $T_c$ and the hump disappears above $T_c$[11,12] because the pseudogap is weaker and the hump is dominated by electron-boson coupling. These spectra are reproduced from the full dataset of the temperature and doping dependence of the antinodal spectra shown in Fig. S1. The result suggests that the rapidly decreasing pseudogap with doping causes this qualitative difference in the temperature dependence of the spectral lineshape between UD92 and OD80 (Figs. S4a and S4b). We note that such insights discussed above cannot be gained only from the spectral lineshape without knowing the exact form of the pseudogap and its interplay with superconductivity. The spectral weight singularity at $T_c$ discussed in the main text provides us with important clues to understand these long standing puzzles.

Additionally, although the spectral-weight singularity suggests that there is no pseudogap at $p = 0.224$ (Fig. 3f), we found that the antinodal gap closes at $T_{AN} \sim 90$ K, which is ~25 K higher than $T_c$ (Figs. 4a and Supplementary Fig. 1). At present, it is unclear whether this gap above $T_c$ in the deeply-overdoped regime is due to fluctuating superconductivity[13,14] or some form of dynamic phase separation.[15] This observation adds a new twist to the rich phase diagram in the overdoped regime.

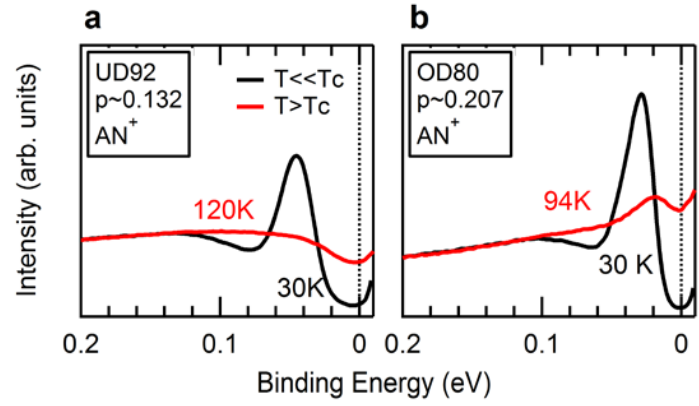

**Fig. S4. Comparison of the temperature dependence of the antinodal spectra between UD92 and OD80. A,b** FD-divided antinodal spectra at $T \ll T_c$ (black) and $T > T_c$ (red) for UD92 and OD80, respectively. These spectra are reproduced from Fig. S1.

*Calculation details*

*Temperature dependence of gaps*

We utilize Ginzburg-Landau theory to derive the temperature dependence of the pseudogap $\Delta_{PG}(T)$ and superconducting gap $\Delta_{SC}(T)$. The Ginzburg-Landau free energy for two coupled order parameters, $\Delta_{SC}$ and $\Delta_{PG}$, is given by

$$\mathcal{F}(\Delta_{SC}, \Delta_{PG}) = A|\Delta_{SC}|^2 + |\Delta_{SC}|^4 + B|\Delta_{PG}|^2 + |\Delta_{PG}|^4 + \gamma|\Delta_{SC}|^2|\Delta_{PG}|^2,$$

where the two order parameters are repulsive if $\gamma > 0$. In order to complete the square in this expression we define

$$|M|^2 = |\Delta_{PG}|^2 + \frac{\gamma}{2}|\Delta_{SC}|^2$$

The expression for F can then be rewritten as

$$\mathcal{F}(\Delta_{SC}, M) = A'|\Delta_{SC}|^2 + B'|M|^2 + C'|\Delta_{SC}|^4 + |M|^4$$

where $A' = A - B\gamma/2$, $B' = B$, and $C' = 1 - \gamma^2/4$. We now minimize F($\Delta_{SC}$,M) with respect to $\Delta_{SC}^*$ and $M^*$ to obtain the temperature dependence of the order parameters:

$$\frac{\partial F}{\partial \Delta_{SC}^*} = \Delta_{SC}(A' + C'|\Delta_{SC}|^2) = 0$$

and

$$\frac{\partial F}{\partial M^*} = M(B' + |M|^2) = 0$$

From this we obtain the solutions $|\Delta_{SC}|^2 = -A'/C'$ and $|M|^2 = -B'$. These equations are satisfied by a BCS form for gap temperature dependence:[16]

$$\Delta_{SC}(T) = \Delta_0 \tanh\left[\alpha\sqrt{\frac{T_c}{T}-1}\right]$$

and

$$M(T) = \Delta_0^{PG} \tanh\left[\alpha\sqrt{\frac{T_c}{T}-1}\right]$$

where $\Delta_0$ and $\Delta_0^{PG}$ are the temperature-independent parts of $A'/C'$ and $B'$, respectively, and $\alpha$ is an external adjustable parameter that controls how fast the gaps turn on. In principal $\alpha$ in the expressions for $\Delta_{SC}(T)$ and $M(T)$ can be different, but we choose them to be the same because this detail is not important for our results.

From the definition of $M$ above we know that $|\Delta_{PG}|^2 = (M - \sqrt{\gamma/2}\Delta_{SC})(M + \sqrt{\gamma/2}\Delta_{SC})$, and since we are interested in a repulsive solution we choose $\Delta_{PG} = M - \sqrt{\gamma/2}\Delta_{SC}$. We can then obtain an expression for the pseudogap temperature dependence:

$$\Delta_{PG}(T) = \Delta_0^{PG} \tanh\left[\alpha\sqrt{\frac{T^*}{T}-1}\right] - \sqrt{\gamma/2}\Delta_0^{SC} \tanh\left[\alpha\sqrt{\frac{T_c}{T}-1}\right]$$

The parameter $\gamma$ controls how much superconductivity suppresses the pseudogap.

The temperature dependence of $\Delta_{SC}(T)$ and $\Delta_{PG}(T)$ are shown in Fig. 3c in the main text. Throughout this work, we take $T^* = 200$ K, $T_c = 90$ K, $\Delta_0^{SC} = 35$ meV, and $\Delta_0^{PG} = 105$ meV.

*Spectral function*

The pseudogap and electron-boson coupling are included in self-energy terms in the electronic Green's function. In the normal state ($T > T_c$) the Green's function is

$$G(\mathbf{k},i\omega_n) = \left[ i\omega_n - \epsilon_{\mathbf{k}} - \Sigma_{PG}(\mathbf{k},i\omega_n) - \Sigma_{ph}(i\omega_n) \right]^{-1}$$

In the superconducting state, we calculate the (1, 1) component of the matrix Green's function in Nambu notation, which is given by

$$G(\mathbf{k},i\omega_n) = \left[ i\omega_n - \epsilon_{\mathbf{k}} - \frac{\Delta_{\mathbf{k}}^2}{i\omega_n + \epsilon_{\mathbf{k}} + \Sigma_{ph}(-i\omega_n)} - \Sigma_{PG}(\mathbf{k},i\omega_n) - \Sigma_{ph}(i\omega_n) \right]^{-1}$$

The spectral function is given by the analytic continuation of $G$:

$$A(\mathbf{k},\omega) = -\frac{1}{\pi} \text{Im } G(\mathbf{k},\omega).$$

*Bandstructure*

We use a bandstructure obtained from a tight-binding fit to ARPES data on Bi2201.[17] The use of this bandstructure for the anti-bonding band in Bi2212 is reasonable because the value of $\mathbf{k}_F$ is similar to that in Bi2201 in the antinodal region. In the simulation, we have assumed a square-lattice Brillouin zone with momentum measured in units where the square-lattice constant *a* has been set equal to unity. This bandstructure has a van Hove singularity at ≈ -38 meV. In order to minimize its effect on the self-energies, for the case of the electron-boson self-energy, the negative frequency self-energy is obtained from the positive frequency part: $\Sigma_{ph}(-\omega) = -\Sigma_{ph}'(\omega) + i\Sigma_{ph}''(\omega)$. In the case of the pseudogap self-energy, and the electron-boson self energy in the superconducting state, a $\mathbf{k}_z$ dispersion of the form $t_\perp \cos \mathbf{k}_z$ is included in the bandstructure, and an integration over $\mathbf{k}_z$ is performed.

*Superconductivity*

We take the superconducting gap to have a *d*-wave form

$$\Delta_{\bm{k}} = (\Delta_0/2)(\cos k_x - \cos k_y),$$

where the gap maximum at temperature $T$ is $\Delta_0 = \Delta_{SC}(T)$, as defined above.

*Pseudogap*

The pseudogap is modeled as a fluctuating density wave (DW) order, using the formalism of Ref. [18]. Treating the pseudogap in this way has successfully reproduced ARPES spectra in the pseudogap regime.[1,2] Within this framework, the self-energy from the fluctuating DW is

$$\Sigma_{PG}(\bm{k}, i\omega_n) = \Delta_{PG}^2 \int d\bm{q}\, \frac{P(\bm{q})}{i\omega_n - \epsilon_{\bm{k}+\bm{q}}},$$

where $\Delta_{PG} = \Delta_{PG}(T)$ as defined above, and $P(\bm{q})$ is a Lorentzian centered at the DW ordering vector $\bm{Q}_{DW}$. The inverse width of the Lorentzian gives the coherence length of the fluctuating DW. We consider a "stripe-like" fluctuating DW with ordering vector $\bm{Q}_{DW} = (\pm 0.2\pi, 0)$:

$$P(\bm{q}) = \frac{1}{\pi}\frac{\Gamma}{(q_x - 0.2\pi)^2 + \Gamma^2} + \frac{1}{\pi}\frac{\Gamma}{(q_x + 0.2\pi)^2 + \Gamma^2}$$

The coherence length is set to $1/\Gamma = 20$.

*Electron-boson coupling*

As discussed in the main text, an inclusion of an electron-boson coupling is necessary to describe temperature evolution of the spectral line shape. In particular, we consider a coupling to phonons. We compute the electron-phonon self-energy using single iteration expressions from Migdal-Eliashberg theory. A phenomenological treatment using the Migdal-Eliashberg self-energy has been shown to describe a number of experimental features of the optimally and overdoped cuprates.[19]

We assume a momentum independent coupling g to dispersionless phonons with frequency Ω = 35 meV. The electron-phonon self-energy is given by

$$\Sigma_{ph}(i\omega_n) = -\frac{g^2}{\beta V} \sum_{\mathbf{k}, ip_n} D_0(ip_n) G_0(\mathbf{k}, i\omega_n - ip_n)$$

Here $D_0$ is the non-interacting phonon Green's function, and $G_0$ is the non-interacting electronic Green's function. In the superconducting state, this self-energy can be written in the form $\Sigma_{ph}(i\omega_n) = i\omega_n(1-Z(i\omega_n)) + \chi(i\omega_n)$, where the imaginary parts of Z and χ are given by

$$\omega Z_2(\omega) = \frac{g^2 \pi}{2N} \sum_k \left[ n_B(\omega) + n_F(E_k) \right] \left[ \delta(\omega + \Omega - E_k) + \delta(\omega - \Omega + E_k) \right]$$
$$+ \left[ n_B(\omega) + n_F(-E_k) \right] \left[ \delta(\omega - \Omega - E_k) + \delta(\omega + \Omega + E_k) \right]$$

and

$$\chi_2(\omega) = \frac{-g^2 \pi}{2N} \sum_k \frac{\varepsilon_k}{E_k} \left[ n_B(\omega) + n_F(E_k) \right] \left[ \delta(\omega + \Omega - E_k) - \delta(\omega - \Omega + E_k) \right]$$
$$+ \frac{\varepsilon_k}{E_k} \left[ n_B(\omega) + n_F(-E_k) \right] \left[ \delta(\omega - \Omega - E_k) - \delta(\omega + \Omega + E_k) \right]$$

Here $n_B$ and $n_F$ are the Bose and Fermi functions, respectively, and $E_k = \sqrt{\varepsilon_k^2 + \Delta_k^2}$. The electron-phonon coupling is chosen consistent with the range of values used in previous studies [20] and defined in terms of the mass renormalization through the real-part of the self-energy, or Z, in the limit $\omega \to 0$ such that the dimensionless coupling λ ~ 0.3.

*Marginal Fermi liquid self-energy*

We include a phenomenological broadening due to a marginal Fermi liquid self energy in the calculation of all spectra:

$$\Sigma_{MFL}(\omega) = \sqrt{\alpha \omega^2 + \beta T^2}$$

We set α = 0.025 and β = 2.5 and measure frequency $\omega$ and temperature $T$ in units of eV.